\documentclass[runningheads]{llncs}
\usepackage[T1]{fontenc}
\usepackage{graphicx}
\usepackage[utf8]{inputenc}
\usepackage{multirow}
\usepackage{pifont}
\usepackage{bm}
\usepackage{listings}
\usepackage{xcolor}
\usepackage{booktabs}
\usepackage{placeins}
\usepackage{hyperref}
\usepackage{stfloats}
\usepackage{array}

\newcommand{\tmark}{$\bm{\triangle}$}

\begin{document}
\title{Fully Automated End-to-End Adversary Emulation from MITRE ATT\&CK Based Cyber Threat Intelligence Using LLMs}
\titlerunning{Fully Automated Adversary Emulation from CTI Using LLMs}
%
%

\author{
Jueon Choi\inst{1}\orcidID{0009-0006-8331-1251}
\and
Seojun Lee\inst{1}\orcidID{0009-0008-7178-2341}
\and
Sanggwon Yun\inst{1}\orcidID{0009-0000-9639-032X}
\and
Kwanghoon Choi\inst{1}\orcidID{0000-0003-3519-3650}
\and
Gunjin Cha\inst{2}\orcidID{0009-0004-7699-6286}
}

\authorrunning{J. Choi et al.}

\institute{
Chonnam National University, Gwangju, Republic of Korea\\
\email{\{jueonchoi0512, tjwns1300, sanggwon, kwanghoon.choi\}@jnu.ac.kr}
\and
Good First Information Technology, Gwangju, Republic of Korea\\
\email{gj.cha@gfict.com}
}

%
%
%
\maketitle              
\begin{abstract}
This paper presents a fully automated end-to-end framework for adversary emulation from MITRE ATT\&CK-aligned CTI reports using LLMs. Unlike prior work, which either executes prewritten playbooks or partially automates playbook generation, our framework unifies playbook generation, execution, and failure recovery in a single workflow. In particular, although AURORA, the most recent prior study, generates playbooks from CTI reports, it still requires partial manual intervention and does not revise playbooks based on execution failures. Our framework generates Caldera playbooks from CTI reports, executes them automatically, and revises failed Abilities through a failure-type-aware recovery mechanism. Evaluated on 11 CTI reports with Claude Sonnet 4.5, GPT-4o, Gemini 2.5 Pro, and Grok 4 Fast, the framework achieved its best results with Claude Sonnet 4.5: 27.3 Abilities per playbook, 84.22\% execution success after revision, and CTI Precision, Recall, and F1 of 73.95\%, 52.48\%, and 60.50\%, respectively. The failure recovery mechanism consistently improved execution success across all evaluated LLM models by 14.59\%p to 17.23\%p. 
On the 10 CTI reports selected from AURORA's dataset, this mechanism further increased the final execution success rate, surpassing that of AURORA, which represents the state-of-the-art adversary emulation system.

\end{abstract}

\section{Introduction}
\label{sec:introduction}
Cyberattack techniques continue to grow in sophistication, and accordingly, defense systems must be equipped with means to realistically reproduce and validate the latest threats\cite{Ahmad2021HowCO}. Such means include attack emulation, BAS (Breach and Attack Simulation)\cite{bas_rapid7}, and Red Teaming\cite{Kovaevi2020RedT}. Adversary emulation\cite{applebaum2016automated} among these is a method that precisely replicates the specific tactics, techniques, and procedures (TTPs) used by a particular threat group (e.g., an APT) according to a predefined scenario, in order to assess in a focused manner whether an organization’s defense systems can detect or block such attacks. 
However, for adversary emulation to truly help improve security systems, organizations need the capability and operational setup to repeatedly reproduce the latest attack techniques on an ongoing basis.

To maximize the effectiveness of adversary emulation, end-to-end automation is required, encompassing scenario design, attack execution, and recovery procedures in the event of errors. To implement such systematic automation, it is necessary to describe an attacker’s actions according to a consistent standard. Existing CTI(Cyber Threat Intelligence) reports often rely on natural-language descriptions that are useful for human analysts, but they are not always sufficiently structured to support direct machine interpretation and automated execution. Therefore, to make attack procedures usable in an automatable form, reports are needed in which attack stages, employed techniques, and attack flows are described in a standardized structure. MITRE ATT\&CK\cite{mitre_attck,strom2018mitre} addresses this need by organizing adversarial TTPs into a knowledge base of tactics, techniques, and procedures. It classifies an attacker's goals (tactics), means of achieving those goals (techniques), and detailed implementation actions (procedures) in a step-by-step manner to strengthen an organization's threat detection and defense systems. By analyzing real-world security incidents and writing CTI reports in accordance with the MITRE ATT\&CK framework, such reports can be used as input for the adversary emulation process.

Existing studies on attack emulation can be broadly classified by the scope of automation they address. Tools such as MITRE Caldera~\cite{caldera}, PurpleSharp~\cite{purplesharp}, and Attack Range~\cite{attackrange} automate the execution of predefined playbooks but leave scenario generation entirely to security experts, requiring repeated manual effort\cite{AkbariGurabi2024Playbook} whenever new threats emerge. AURORA~\cite{aurora} extends automation to playbook generation by using an LLM to extract tactics and techniques from CTI reports and constructing attack chains through symbolic planning that models preconditions and effects. However, AURORA still requires manual intervention at certain stages, such as serving and executing payloads. Furthermore, the system lacks a mechanism for diagnosing or repairing failures during execution;  when a command fails, it simply proceeds to the next step in the chain. Because the same command can yield different outcomes depending on subtle environmental differences, such unhandled failures invalidate the planned state transitions and weaken the execution validity of the entire attack plan.

Based on an analysis of the strengths and limitations of prior studies, three key elements that automated attack emulation should commonly provide were identified: playbook generation, playbook execution, and failure recovery.
Ultimately, these three elements reflect the practical workflow of a security analyst: designing an attack scenario (\textit{playbook generation}), executing it against a target (\textit{playbook execution}), and resolving runtime failures (\textit{failure recovery}). Because the absence of any single element necessitates manual intervention, integrating all three is essential to achieve fully automated, end-to-end adversary emulation.

\begin{table}[ht]
\centering
\caption{Comparison of the Proposed System with Prior Studies}
\label{tab:framework-comparison}
\scriptsize
\begin{tabular}{|m{2.8cm}|c|c|c|}
\hline
Approach & Auto Generation & Auto Execution & Failure Recovery \\ \hline
Caldera, PurpleSharp, Attack Range, Perry & X & O & X \\ \hline
AURORA & O & \tmark & X \\ \hline
Ours (Proposed) & O & O & O \\ \hline
\end{tabular}
\end{table}

A comparative analysis of prior studies with respect to these three elements is summarized in Table~\ref{tab:framework-comparison}. As shown in the table, among existing studies on automated attack emulation, no approach has yet been proposed that automatically generates playbooks, executes them automatically, and revises them by incorporating feedback from execution results, thereby ultimately reducing time and cost while maximizing the attack success rate.

\begin{figure}[ht]
    \centering
    \includegraphics[width=0.8\linewidth]{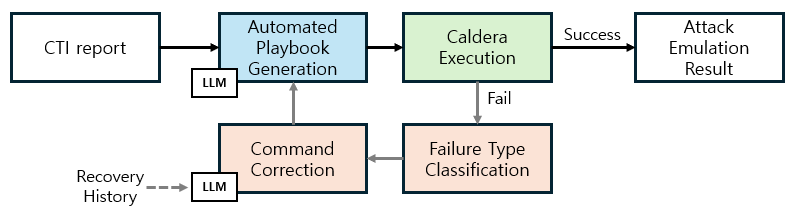}
    \caption{System Overview}
    \label{fig:system-overview}
\end{figure}

In this study, we present a fully automated end-to-end framework for adversary emulation from MITRE ATT\&CK-aligned CTI reports using LLMs. Rather than improving only a single stage of the pipeline, our framework unifies playbook generation, execution, and failure recovery into one automated workflow, as illustrated in Figure~\ref{fig:system-overview}. The system uses an LLM to extract attack goals and methods from natural-language CTI reports and combines them with an environment specification to generate executable commands and their dependencies. The generated results are converted into MITRE Caldera Abilities and assembled into an Adversary--which we regard as a playbook in this study--that is automatically uploaded and executed on the MITRE Caldera platform without human intervention.

We evaluated the framework with four LLMs--Claude Sonnet 4.5~\cite{anthropic2025claude}, GPT-4o~\cite{openai2024gpt4o}, Gemini 2.5 Pro~\cite{google2025gemini}, and Grok 4 Fast~\cite{xai2025grok}. Among them, Claude Sonnet 4.5 achieved the best performance: 94.91\% ATT\&CK conformance, an average of 27.3 Abilities per playbook, an execution success rate of 84.22\%, and a CTI F1-score of 60.50\%. 

To assess the competitiveness of our system, we conducted a comparative evaluation using 10 CTI reports from the dataset of AURORA, which represents
the current state of the art. We compared the final executable attack chains generated by the two systems using the same evaluation metrics.
Our system outperformed AURORA in both CTI F1-score (30.57\% vs. 26.07\%) and execution success rate (65.17\% vs. 60.72\%), while additionally providing full end-to-end automation and failure-aware revision.

A central component of the framework is failure-type-aware recovery. The mechanism classifies execution failures into four types--syntax error, environment mismatch, dependency error, and timeout--and applies a tailored recovery strategy to each. All four LLMs achieved consistent improvements of 14.59\%p to 17.23\%p after recovery. 
%
In particular, in the AURORA comparison, our system's initial playbooks achieved only 32.90\% execution success, but after failure recovery the rate rose by 32.27\%p to 65.17\%, demonstrating that the recovery mechanism remains effective even when applied to CTI reports from a different source and environment.

Finally, the framework also provides substantial efficiency gains: generating and revising a playbook required only 197.39 seconds and \$0.35 in LLM token cost per report on average, whereas the ATT\&CK Evaluations Library~\cite{citd_eval} has published only 14 manually constructed attack scenarios over six years (2019--2025), underscoring the scalability advantage of our automated approach.
More details of the experimental results are provided in Section~\ref{sec:evaluation}.

Although prior studies \cite{huang2024selfcorrection,selfdebugging,cycle,shinn2023reflexion,madaan2023selfrefine,pan2024selfcorrection,olausson2024selfrepair} have explored feedback-based correction for general code generation, they have not addressed the attack-simulation domain. 
Our novelty lies in being the first to demonstrate an end-to-end adversary emulation system that automatically performs playbook generation, execution, and failure recovery. Because attack simulation involves heterogeneous and sometimes unrecoverable failures  
in remote target environments that are not fully under the system's control, our recovery mechanism is domain-specific rather than a straightforward reuse of existing ideas.

The main contributions of this study are summarized as follows.
\begin{itemize}
\item
We present a fully automated end-to-end adversary emulation framework that integrates playbook generation, execution, and failure recovery into a single workflow from CTI reports to attack playbooks with high execution success.

\item
We propose a failure recovery mechanism that identifies four execution failure types and revises playbooks with type-specific recovery strategies and prior revision histories, which is essential to practical end-to-end adversary emulation.

\item
We evaluated four commercial LLMs--Claude Sonnet 4.5, GPT-4o, Gemini 2.5 Pro, and Grok 4 Fast--and showed that each model has a distinct area of strength.
\end{itemize}

The remainder of this paper is organized as follows.
Section~\ref{sec:related_work} reviews prior studies. 
Section~\ref{sec:system} describes the proposed system. 
Section~\ref{sec:evaluation} introduces research questions and presents the corresponding experimental results. 
Section~\ref{sec:discussion} presents the results of our comparison with AURORA. Finally, Section~\ref{sec:conclusion} concludes the paper.

\section{Related Work}
\label{sec:related_work}
This section reviews prior work most relevant to our study from three perspectives: cyber attack emulation, AURORA, and failure recovery. 

\subsection{Cyber Attack Emulation}
\label{sec:related_emulation}
Prior studies on cyber attack emulation span a wide spectrum of automation~\cite{aurora}. 
%
At one end, numerous studies have actively extracted attacker behaviors from CTI reports and mapped them to structured representation frameworks such as MITRE ATT\&CK. For example, recent work has leveraged LLMs to automatically extract entities and relationships in STIX---a standard for structured cyber threat representation--- from unstructured CTI text~\cite{lekssays2025azerg}, aiming to structure threat intelligence and reduce the manual effort required of security analysts. However, these approaches remain at the level of abstract modeling and do not produce executable artifacts.
At the other end, tools such as Atomic Red Team~\cite{atomicredteam}, Metasploit~\cite{metasploit}, and LOLBAS~\cite{LOLBAS} provide executable means for individual techniques, and frameworks such as MITRE Caldera~\cite{caldera}, PurpleSharp~\cite{purplesharp}, and Splunk Attack Range~\cite{attackrange} orchestrate them into multi-stage playbooks. 
However, these tools require playbooks to be manually authored by security experts: even the MITRE Engenuity ATT\&CK Evaluations~\cite{citd_eval} have published only 14 executable scenarios over six years (2019--2025), underscoring the scalability bottleneck of manual approaches. 
More recently, Perry et al. [25] proposed a high-level abstraction layer that reduces the implementation effort for attack and deception strategies, but the strategies themselves must still be manually specified by the operator.

\subsection{AURORA}
\label{sec:related_aurora}
AURORA~\cite{aurora} is an end-to-end system for automatically generating multi-stage attack scenarios from CTI reports, and is the prior work most closely related to our study. AURORA extracts ATT\&CK techniques from CTI reports using an LLM-based report analyzer, generates attack chains that preserve causal relationships among attack steps using a PDDL (Planning Domain Definition Language)\cite{mcdermott1998pddl}-based emulation planner.
The authors demonstrated its effectiveness by applying it to 250 CTI reports and generating corresponding attack chains.

AURORA has two main limitations. First, it still depends on human intervention at certain stages, such as serving and executing payloads. For example, the generated \texttt{attack\_chain.py} may include \texttt{[MANUAL ACTION REQUIRED]} blocks.
%
Second, AURORA does not revise failed commands based on execution feedback, which can leave certain attack stages untested and give organizations an incomplete view of their defensive posture.

\subsection{Failure Recovery}
Research on LLM self-correction has mainly been developed in general code generation~\cite{huang2024selfcorrection}. Representative studies have shown that LLMs can revise their code based on execution feedback~\cite{selfdebugging,cycle}, that iterative self-feedback loops improve LLM outputs across diverse tasks~\cite{shinn2023reflexion,madaan2023selfrefine}, and that feedback from external tools or execution environments is generally more effective than internally generated feedback~\cite{olausson2024selfrepair,pan2024selfcorrection}. However, all of the above studies focus on general code generation, and no prior work has systematically applied failure-type taxonomy and recovery strategies specialized for the attack-simulation domain.

The attack-simulation domain differs fundamentally from general code generation in three ways: heterogeneous failure types, the existence of unrecoverable failures, and limited control over the execution environment. Unlike general code errors, attack-command failures arise from diverse causes and often require different recovery strategies. Some failures are inherently unrecoverable, and attack commands are executed in remote target environments whose state is not fully controllable.

Motivated by the finding of Pan et al.~\cite{pan2024selfcorrection} that leveraging feedback from external sources such as execution results is more effective than relying on the LLM's own self-assessment, 
we incorporate execution feedback from Caldera into a failure recovery mechanism that combines failure-type classification, type-specific recovery strategies, and the use of prior recovery history.

\section{System Design}
\label{sec:system}
\begin{figure}[ht]
    \centering
    \includegraphics[width=0.7\linewidth]{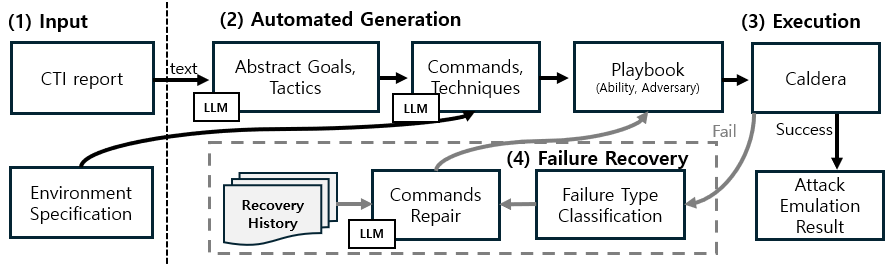}
    \caption{System Architecture}
    \label{fig:system-architecture}
\end{figure}

Our system is an automated framework that uses an LLM to transform CTI reports into Caldera playbooks and automatically recovers from execution failures through a recovery mechanism. As shown in Figure~\ref{fig:system-architecture}, the overall system consists of four functional layers: CTI and Environment Input, automated playbook generation, playbook execution, and failure recovery.

\subsection{CTI and Environment Input Layer}
\subsubsection{CTI report.}
The input to our system is a collection of 11 high-quality CTI reports, each of which contains multiple TTPs, published by the Korea Internet \& Security Agency (KISA)\cite{kisa_ttp_reports}. Published between 2020 and 2024, these reports are structured based on the MITRE ATT\&CK framework and cover a wide range of cyber threats, including ransomware, APT attacks, surveillance-oriented attacks, and web application attacks. Each report is based on a real intrusion incident and provides detailed descriptions of the attacker's TTPs. Details of the dataset are presented in Section~\ref{sec:cti-dataset}.

Each report is organized hierarchically by tactics (e.g., \textit{Initial Access}), with specific techniques enumerated under each tactic (e.g., \textit{Valid Accounts}, \textit{Exploit Public-Facing Application}). For each technique, the report provides concrete operational details such as web request traces, file paths, and execution commands, along with corresponding mitigation strategies. This combination of ATT\&CK-aligned structure and operational-level detail makes the reports particularly well-suited as input for automated attack emulation.

However, the reports contain environment-specific details from the real incident, such as IP addresses, account information, and file paths. These details must be abstracted so that the scenario can be reused across different environments.

\subsubsection{Environment Specification.}
\label{sec:env_spec}
The environment specification is a document written by a security operator that provides essential information for concretizing abstract attack goals in the target environment. It includes network and system information, such as the IP addresses of target hosts, the structure of the internal network, operating system versions, installed software, and user account information. The LLM uses this specification to translate environment-dependent information in the CTI report into concrete commands for the test environment. 

\subsection{LLM-Based Playbook Generation Layer}
The LLM-based playbook generation layer transforms a CTI report into a playbook executable in Caldera through three stages. 

\subsubsection{Step 1: Abstract Attack Flow Generation.}
Step 1 extracts text from the CTI report in PDF format and converts it into a structured representation separated by page. Because the reports average 41.5 pages, page-level segmentation makes the input more manageable for subsequent LLM processing. The system then iterates through the extracted text in chunks, progressively identifying attack goals from the natural-language narrative. Each attack goal consists of an attack objective, a MITRE ATT\&CK tactic, and a description of the method used to achieve the objective. The resulting collection of attack goals forms an environment-independent abstract attack flow.

\subsubsection{Step 2: Concrete Attack Flow Generation.}
In Step 2, the system provides the abstract attack flow from Step 1 and the environment specification to the LLM, which converts each attack goal into one or more nodes. Each node includes an executable command, a MITRE ATT\&CK Technique ID, and environment-specific information such as IP addresses, paths, and credentials. Logical prerequisite relationships among nodes are explicitly represented as edges, and an execution order is determined so that the resulting attack flow can be executed in that order.

\subsubsection{Step 3: Caldera Ability and Adversary Generation.}
Step 3 converts each node from Step 2 into a Caldera Ability containing the node's command, tactic, and technique ID. The Abilities are then assembled into an Adversary by ordering them according to the execution order determined in Step 2. The final result is stored in \texttt{abilities.yml} and \texttt{adversaries.yml}, ready to be uploaded to Caldera for automated execution.

\subsection{Caldera Execution Layer}
MITRE Caldera is an open-source platform for automating ATT\&CK-based attack emulation~\cite{caldera,applebaum2016automated}. An Ability is an execution unit that implements a specific ATT\&CK technique as operating-system-specific commands, and an Adversary is an attack profile that arranges multiple Abilities in a logical order. An Agent resides on a target host, receives Abilities from the Server, executes them, and returns the results, including exit code, stdout, and stderr. The Caldera execution layer uploads the generated Abilities and Adversary from the previous layer, executes them via the Agent, and passes the collected results to the Failure Recovery Layer.

\begin{figure}[ht]
\centering
\includegraphics[width=0.9\linewidth, height=0.4\textheight]{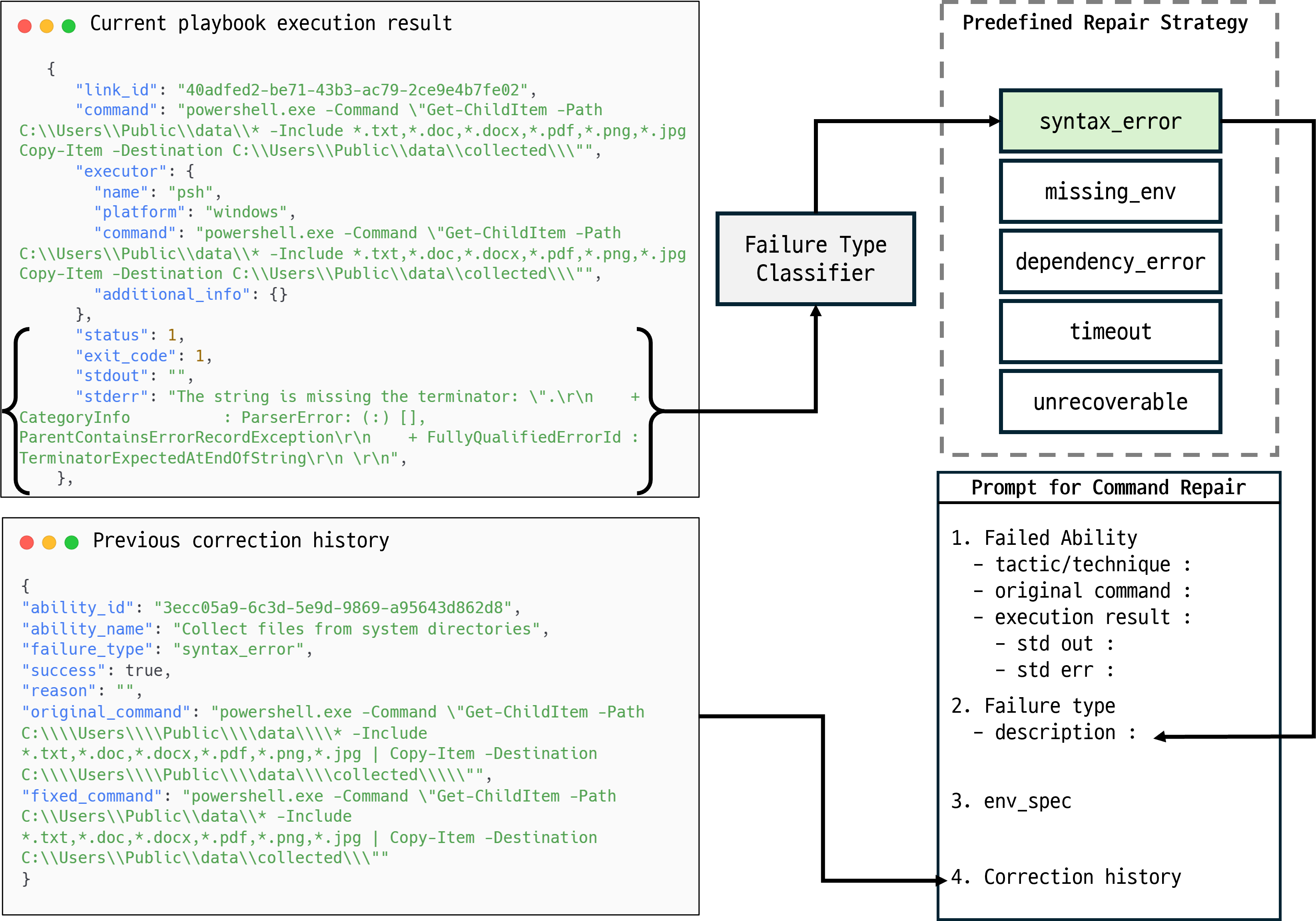}
\caption{Classifying failure types from execution results and constructing a recovery prompt with the corresponding strategy and revision history.}
\label{fig:self-correction}
\end{figure}

\subsection{Failure Recovery Layer}
In the attack-simulation domain, the same command may succeed or fail depending on subtle differences in the target environment, such as its OS version, security policy, and network configuration, and a failure in an earlier stage can cascade and invalidate all subsequent stages. The Failure Recovery Layer addresses this by classifying failures into predefined types and applying tailored recovery strategies, as shown in Figure~\ref{fig:self-correction}.

\begin{table}[ht]
\centering
\caption{Failure type classification patterns}
\label{tab:failure-patterns}
\scriptsize
\setlength{\tabcolsep}{2pt}
\begin{tabular}{|l|p{8.5cm}|}
\hline
\textbf{Type} & \textbf{Patterns} \\ \hline
SYNTAX\_ERROR &
parsererror ~$\mid$~
unexpected token ~$\mid$~
At line:\textbackslash d+ ~$\mid$~
is not a valid statement separator ~$\mid$~
invalid\textbackslash s+(key\textbackslash s+name|parameter) ~$\mid$~
positional parameter ~$\mid$~
cannot be found that accepts ~$\mid$~
incorrectly formatted ~$\mid$~
is not recognized as the name of a cmdlet ~$\mid$~
service name is invalid 
\\ \hline
DEPENDENCY\_ERROR &
access\textbackslash s*(is\textbackslash s*)?denied ~$\mid$~
requires?\textbackslash s+elevation ~$\mid$~
privilege ~$\mid$~
unauthori[sz]ed ~$\mid$~
run\textbackslash s+as\textbackslash s+administrator ~$\mid$~
(open|start|create)service \textbackslash s+failed ~$\mid$~
winrm\textbackslash s+client \textbackslash s+cannot\textbackslash s+process 
\\ \hline
MISSING\_ENV &
cannot find (path|drive) ~$\mid$~
does not exist ~$\mid$~
network name cannot be found ~$\mid$~
unable to connect ~$\mid$~
system error\textbackslash s+(53|67|86|1355|6118) 
\\ \hline
TIMEOUT &
timeout\textbackslash s*reached ~$\mid$~
process\textbackslash s*(was\textbackslash s*)?killed ~$\mid$~
did not respond.+in a timely fashion ~$\mid$~
Exit code: 124 
\\ \hline
\end{tabular}
\end{table}

\textbf{Domain-Specific Failure Classification.}
The system extracts error logs from the execution result of each Ability and classifies failure types using predefined regular-expression patterns derived from error patterns actually observed in the attack-simulation environment. The classification consists of four types: \texttt{syntax\_error} (command syntax errors such as Escaping and string parsing errors), \texttt{dependency\_error} (privilege and cascading failure issues), \texttt{missing\_env} (absence of required environmental resources such as file paths or network endpoints), and \texttt{timeout} (execution exceeding the time limit). Failures that do not match any of these types are excluded from the recovery loop.
Table~\ref{tab:failure-patterns} summarizes the regular-expression patterns used to detect and classify each failure type.

\begin{table}[ht]
\centering
\caption{Failure type definitions and their specifications}
\label{tab:failure-spec}
\scriptsize
\setlength{\tabcolsep}{2pt}
\begin{tabular}{|l|p{9.5cm}|}
\hline
\textbf{Failure Type} & \textbf{Specification} \\ \hline
\texttt{SYNTAX\_ERROR} &
  The command syntax is invalid and cannot be parsed or executed.
  This includes incorrect parameter formats, missing arguments,
  incompatible command usage, and string/path format errors. \\ \hline
\texttt{DEPENDENCY\_ERROR} &
  The prerequisites required for command execution are not met.
  This includes insufficient permissions, cascading failures due to prior task failures,
  authentication failures, remote connection configuration issues, and service access denials. \\ \hline
\texttt{MISSING\_ENV} &
  The file, path, network resource, or service referenced by the command does not exist
  in the current environment. This includes incorrect IP/URL, non-existent file paths,
  and network connection failures. \\ \hline
\texttt{TIMEOUT} &
  The command execution did not complete within the specified time limit.
  This includes long-running operations, timeout while waiting for response,
  and service response delays. \\ \hline
\end{tabular}
\end{table}

\textbf{Recovery with Accumulated Revision History.}
For each recovery attempt, the system accumulates failure logs and revision records from all previous attempts and includes them in the prompt, along with the original command and the environment specification.
To guide the LLM toward a targeted repair based on the classified failure type, the prompt also includes the corresponding definition from Table~\ref{tab:failure-spec}, which describes the scope and characteristics of each type. 
This design both discourages repeated errors across retries and directs the LLM toward a failure-type-specific repair.

Figure~\ref{fig:self-correction} illustrates the overall recovery process.
The failure recovery process is repeated up to three times, reflecting a trade-off between cost and recovery effectiveness, as recoverable failures were empirically resolved within this limit. The process terminates when all Abilities execute successfully, when the retry limit is reached, or when no further failures are classified as recoverable types.

\section{Evaluation}
\label{sec:evaluation}
In this section, we evaluate the proposed system through a series of experiments. We first describe the experimental setup and then discuss the results in relation to our four research questions.

\begin{figure}[ht]
\centering
\includegraphics[width=0.4\linewidth]{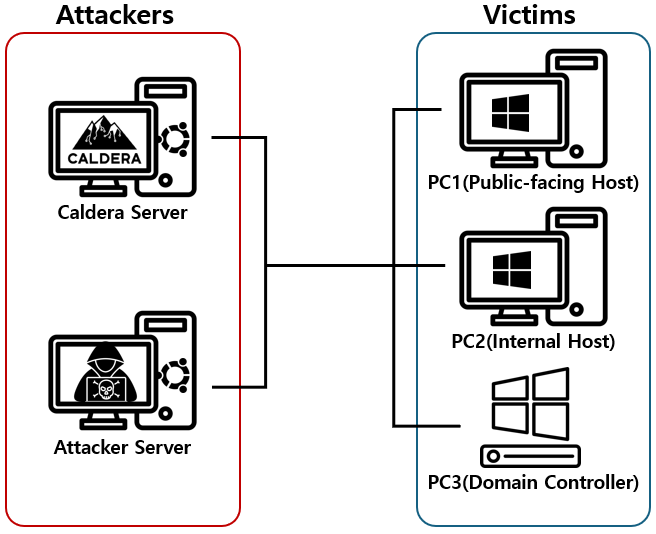}
\caption{Network and System Configuration of the Experimental Environment}
\label{fig:network}
\end{figure}

\subsection{Test Environment}
\subsubsection{Experimental Setup}
Figure~\ref{fig:network} shows the experimental environment, which consists of three system roles: \textit{PC1}, an externally exposed initial-access point; \textit{PC2}, an internal host reachable only via lateral movement; and \textit{PC3}, a central enterprise management host, such as an Active Directory domain controller. This configuration covers the essential stages of enterprise internal-network intrusion---initial compromise, lateral movement, and domain control. The entire network was isolated from the external Internet, and security solutions such as Windows Defender were disabled to ensure that the evaluation focused solely on the quality of generated playbooks.

Specifically, the experimental setup is as follows.
A Caldera server (Caldera 5.3.0 on Ubuntu 20.04 LTS) manages playbook upload, execution, and result collection, while an attacker server provides payload delivery and data exfiltration endpoints. A Caldera agent on PC1 receives commands from the server, executes them, and returns results. 
PC1 and PC2 run Windows 10 Pro, while PC3 runs Windows Server 2019.
Table~\ref{tab:pc-usage} summarizes the system roles required by each of the 11 CTI reports.

\begin{table}[ht]
\centering
\caption{Experimental-environment system roles used by each CTI report (TTPs\#1--\#11)}
\label{tab:pc-usage}
\scriptsize
\setlength{\tabcolsep}{2pt}
\begin{tabular}{|l|l|c|c|c|c|c|c|c|c|c|c|c|}
\hline
\textbf{Role} & \textbf{OS} & \textbf{\#1} & \textbf{\#2} & \textbf{\#3} & \textbf{\#4} & \textbf{\#5} & \textbf{\#6} & \textbf{\#7} & \textbf{\#8} & \textbf{\#9} & \textbf{\#10} & \textbf{\#11} \\ \hline
PC1 (Public-facing host) & Win 10 Pro & O & O & O & O & O & O & O & O & O & O & O \\ \hline
PC2 (Internal host) & Win 10 Pro &O &  &  &  & O &  & O & O &  & O & O \\ \hline
PC3 (Domain controller) & Win Srv 2019 &  &  &  &  & O &  &  & O &  &  &  \\ \hline
\end{tabular}
\end{table}

We also prepared environment-specific prerequisites for each report, such as IIS/ASP for web application attack scenarios and SMB shares for file-sharing attack scenarios.

\subsubsection{CTI Report Dataset}
\label{sec:cti-dataset}

Table~\ref{tab:cti-report-dataset} shows the 11 KISA TTPs reports~\cite{kisa_ttp_reports} used in our experiments. These reports systematically describe attack techniques identified through investigations of real security incidents, structured based on the MITRE ATT\&CK framework. Published between 2020 and 2024, they cover diverse cyber threats including ransomware, APT attacks, surveillance-oriented attacks, and web application attacks. In total, the dataset contains 456 pages and 339,244 words, averaging 41.5 pages and 30,840 words per report.

\begin{table}[ht]
\centering
\caption{Our CTI report dataset}
\label{tab:cti-report-dataset}
\scriptsize
\setlength{\tabcolsep}{2pt}
\begin{tabular}{|l|p{3.4cm}|p{2.8cm}|p{1.8cm}|c|c|c|}
\hline
\textbf{ID} & \textbf{Report Title} & \textbf{Incident Type} & \textbf{Attack Obj.} & \textbf{Comp.} & \textbf{Pg.} & \textbf{Word} \\ \hline
\#1  & Internal Network Compromise through a Website & Web Vulnerability \& Internal Compromise & Exfil. T1041 & Advanced & 32 & 21,397 \\ \hline
\#2  & Info. Collection Attack Infra. through Spear Phishing & Social Engineering \& Phishing & Exfil. T1041 & Expert & 79 & 59,104 \\ \hline
\#3  & The Attacker's Malware Utilization Strategy & Malware Multi-Stage Intrusion & Exfil. T1041 & Advanced & 27 & 15,896 \\ \hline
\#4  & Phishing Target Recon. and Attack Resource Dev. & Reconnaissance \& Resource Development & Exfil. T1041 & Advanced & 61 & 31,630 \\ \hline
\#5  & Attack Patterns Threatening AD Environments & Directory Service Attack & Exfil. T1041, Impact T1486 & Expert & 52 & 39,003 \\ \hline
\#6  & Targeted Watering-Hole Attack Strategy & Website Compromise \& Distribution & Exfil. T1041 & Medium & 29 & 20,506 \\ \hline
\#7  & Internal Network Movement Strategy Using SMB Admin Share & Lateral Movement in Internal Networks & Impact T1486 & Advanced & 32 & 19,373 \\ \hline
\#8  & Op. GWISIN -- Customized Ransomware Attack Strategy & Targeted Ransomware & Exfil. T1041, Impact T1486 & Expert & 35 & 37,872 \\ \hline
\#9  & Attack Strategies for Monitoring an Individual's Daily Life & Spyware \& Personal Surveillance & Exfil. T1041 & Medium & 36 & 27,686 \\ \hline
\#10 & Op. GoldGoblin -- Selective Intrusion Based on a Zero-Day Vuln. & Zero-Day \& APT & Exfil. T1041 & Expert & 40 & 32,637 \\ \hline
\#11 & Op. An Octopus -- Attack on a Centralized Mgmt. Solution & Supply Chain \& Management Systems & Exfil. T1041 & Expert & 33 & 34,140 \\ \hline
\end{tabular}
\end{table}

The final attack goal for each report was determined by the ultimate objective of the attacker as explicitly described in the corresponding KISA report. These goals fall into two categories:
Exfiltration T1041 (\textit{Exfiltration Over C2 Channel}), which appears in 10 of the 11 reports, and Impact T1486 (\textit{Data Encrypted for Impact}), which appears in TTPs\#5, TTPs\#7, and TTPs\#8. We also classify the reports into three complexity levels based on the number of \texttt{attack\_goals} in the Step 1 output: Medium ($\leq 23$), Advanced ($23 < \texttt{attack\_goals} \leq 35$), and Expert ($35 < \texttt{attack\_goals}$).

\subsubsection{Security Verification and Attack Goal Achievement.}
To verify that the experimental environment represents a sufficiently secured system, we executed assessment scripts derived from the KISA Detailed Guide for Technical Vulnerability Analysis and Assessment~\cite{kisa2021checklist} to all three hosts. Specifically, 19 PC vulnerability inspection items (PC-01--PC-19) were evaluated on PC1 and PC2, and 82 Windows server inspection items (W-01--W-82) were evaluated on PC3. The resulting pass rates were 89.95\% for PC1, 89.47\% for PC2, and 93.90\% for PC3, yielding an overall average of 91.11\%.
The failed items on PC1 and PC2 were limited to the latest security patch (PC-09) and firewall activation (PC-11), while on PC3, five items were assessed as non-compliant: administrator account renaming (W-01), latest service pack (W-30), latest security patch (W-32), antivirus installation (W-33), and screen saver configuration (W-36). Because these are either deliberate experiment prerequisites or characteristics of the virtualized environment, we consider the experimental environment to be adequately secured for evaluation purposes.

Despite this secured configuration, the playbooks generated from all 11 CTI reports successfully achieved their designated final attack goals as listed in Table~\ref{tab:cti-report-dataset}. In every case, a valid Ability corresponding to the final attack goal's tactic and technique was both generated and successfully executed, confirming that the proposed framework can carry out complete attack scenarios even against environments with high checklist compliance.

\subsection{Results}
For each report--model pair, we conduct five runs and report the average to account for the probabilistic characteristics of LLM outputs.

\medskip
\noindent\textbf{RQ1: How faithfully do the generated playbooks reproduce the ATT\&CK techniques in the original CTI reports?}

We evaluate the fidelity of generated playbooks using three metrics: \textit{ATT\&CK Validity}, the proportion of generated Abilities whose Technique IDs exist in MITRE ATT\&CK (v15.1, Enterprise)~\cite{mitre_attck,mitre_cti}; \textit{CTI Precision}, the proportion of unique techniques in the generated playbook that also appear in the CTI report; and \textit{CTI Recall}, the proportion of techniques in the CTI report that are reproduced in the generated playbook.

Table~\ref{tab:rq3} presents the results for Claude Sonnet 4.5, which outperformed the other models. ATT\&CK Validity averaged 94.91\%, indicating that most generated techniques correspond to valid MITRE ATT\&CK techniques. CTI Precision averaged 73.95\%, showing that a substantial portion of the generated techniques matched those in the original reports. CTI Recall averaged 52.48\%, indicating that the generated playbooks covered more than half of the techniques described in the reports. 
Together, these results correspond to a CTI F1-score of 60.5\%.
In Section~\ref{sec:discussion}, we present a comparative experiment showing that our system achieves reasonably higher CTI F1-score than AURORA.

\begin{table}[ht]
\centering
\caption{Per-Report MITRE ATT\&CK Validity, CTI Precision, CTI Recall,  CTI F1-score (Claude Sonnet 4.5)}
\label{tab:rq3}
\scriptsize
\setlength{\tabcolsep}{2pt}
\begin{tabular}{|l|c|c|c|c|c|}
\hline
\textbf{ID} & \textbf{Validity(\%)} & \textbf{Prec.(\%)} & \textbf{Recall(\%)} & \textbf{CTI F1(\%)} & \textbf{Complx.} \\ \hline
TTPs\#1  & 98.69  & 92.56  & 61.03 & 73.56 & Advanced \\ \hline
TTPs\#2  & 96.55  & 94.69  & 41.54 & 57.75 & Expert   \\ \hline
TTPs\#3  & 100.00 & 85.18  & 69.56 & 76.58 & Advanced \\ \hline
TTPs\#4  & 85.37  & 66.97  & 40.57 & 50.53 & Advanced \\ \hline
TTPs\#5  & 99.31  & 81.83  & 44.50 & 57.65 & Expert   \\ \hline
TTPs\#6  & 88.02  & 68.69  & 53.04 & 59.86 & Medium   \\ \hline
TTPs\#7  & 93.04  & 71.59  & 52.59 & 60.64 & Advanced \\ \hline
TTPs\#8  & 100.00 & 59.81  & 36.92 & 45.66 & Expert   \\ \hline
TTPs\#9  & 93.96  & 84.63  & 64.55 & 73.24 & Medium   \\ \hline
TTPs\#10 & 100.00 & 53.78  & 52.94 & 53.36 & Expert   \\ \hline
TTPs\#11 & 89.03  & 53.77  & 60.00 & 56.71 & Expert   \\ \hline
\textbf{Average} & \textbf{94.91} & \textbf{73.95} & \textbf{52.48} & \textbf{60.50} & - \\ \hline
\end{tabular}
\end{table}

As a representative case of high Precision but low Recall, we examined the TTPs\#2 report. In this case,
19 techniques were common to both the generated playbook and the report, 2 appeared only in the playbook (Precision: 0.91), and 35 appeared only in the report (Recall: 0.35). The playbook faithfully reproduced the core attack flow, but advanced techniques such as \texttt{Web Shell(T1505.003)} and \texttt{Credentials in Files(T1552.001)} were absent.

To understand why Recall remained low in this representative case, we further analyzed the unreproduced techniques using Claude Sonnet 4.5 on TTPs\#2 as a representative case. 
One interesting phenomenon we observed was that some ATT\&CK techniques were not generated as separate explicit steps, but were instead functionally absorbed into other implemented techniques.
For example, T1569.002 (Service Execution) was not realized as a separate step, but instead incorporated into T1543.003 (Windows Service) as part of a single create-and-start sequence. Likewise, T1074 (Data Staged) was subsumed into the data compression step, T1560 (Data Compressed). This technique-absorption pattern is consistent with the knowledge overshadowing phenomenon~\cite{zhang2025overshadowing}, where dominant knowledge suppresses less salient knowledge during generation, causing minor techniques to lose their semantic distinctiveness and be folded into more prominent ones.

\medskip
\noindent\textbf{RQ2: What is the execution success rate of the generated playbooks, and how much does the failure recovery mechanism improve it?}

Let $N$ denote the total number of Abilities, $M_1$ the number successfully executed on the initial attempt, and $M_2$ the number successfully executed after applying the failure recovery mechanism. We measure four metrics: initial execution success rate ($M_1/N$), final execution success rate ($M_2/N$), improvement rate ($(M_2 - M_1)/N$), and recovery rate ($(M_2 - M_1)/(N - M_1)$). Note that the recovery rate represents the proportion of initially failed Abilities that were successfully recovered.

\begin{table}[ht]
\centering
\caption{Per-Report Execution Success Rates Before and After Failure Recovery (Claude Sonnet 4.5)}
\label{tab:rq2-main}
\scriptsize
\setlength{\tabcolsep}{2pt}
\begin{tabular}{|l|c|c|c|c|c|c|}
\hline
\textbf{ID} & \textbf{Len.} & \textbf{Initial Exec.(\%)} & \textbf{Final Exec.(\%)} & \textbf{Improv.(pp)} & \textbf{Mod.(\%)} & \textbf{Retries} \\ \hline
TTPs\#1  & 33.6 & 52.73 & 70.82 & +18.09 & 49.20 & 3.0 \\ \hline
TTPs\#2  & 29.4 & 73.46 & 96.05 & +22.59 & 27.23 & 2.6 \\ \hline
TTPs\#3  & 23.0 & 70.26 & 91.26 & +21.00 & 29.74 & 2.8 \\ \hline
TTPs\#4  & 30.4 & 63.59 & 81.21 & +17.62 & 36.41 & 3.0 \\ \hline
TTPs\#5  & 27.2 & 64.20 & 77.87 & +13.67 & 36.49 & 3.0 \\ \hline
TTPs\#6  & 24.2 & 74.08 & 85.20 & +11.13 & 25.92 & 2.8 \\ \hline
TTPs\#7  & 28.2 & 81.88 & 88.47 & +6.59  & 18.12 & 3.0 \\ \hline
TTPs\#8  & 33.8 & 52.51 & 86.28 & +33.77 & 49.43 & 3.0 \\ \hline
TTPs\#9  & 23.4 & 82.08 & 88.79 & +6.71  & 17.92 & 2.8 \\ \hline
TTPs\#10 & 22.6 & 85.88 & 92.00 & +6.11  & 14.12 & 2.4 \\ \hline
TTPs\#11 & 24.0 & 65.32 & 68.52 & +3.20  & 34.68 & 3.0 \\ \hline
\textbf{Average} & \textbf{27.3} & \textbf{69.63} & \textbf{84.22} & \textbf{+14.59} & \textbf{30.84} & \textbf{2.85} \\ \hline
\end{tabular}
\end{table}

Table~\ref{tab:rq2-main} shows the per-report results using Claude Sonnet 4.5, the best-performing model. On average, each playbook contained 27.3 Abilities, of which 69.63\% succeeded on the initial attempt. After applying the failure recovery mechanism for up to three retries, the success rate increased to 84.22\%, an improvement of 14.59 percentage points(pp).

\begin{table}[ht]
\centering
\caption{Recovery Rate by Failure Type (Claude Sonnet 4.5, 11 Reports)}
\label{tab:rq2-failure}
\scriptsize
\begin{tabular}{|l|r|r|r|}
\hline
\textbf{Failure Type} & \textbf{\# of Failures} & \textbf{\# of Recovered} & \textbf{Recovery Rate(\%)} \\ \hline
syntax\_error      & 245 & 101 & 41.22 \\ \hline
dependency\_error  & 117 &  32 & 27.35 \\ \hline
missing\_env       &  31 &  19 & 61.29 \\ \hline
timeout            &  26 &   4 & 15.38 \\ \hline
unrecoverable      &  57 &   0 &  0.00 \\ \hline
\textbf{Total} & \textbf{476} & \textbf{156} & \textbf{32.77} \\ \hline
\end{tabular}
\end{table}

Table~\ref{tab:rq2-failure} breaks down the 476 total failure cases by type. Notably, \texttt{syntax\_error} (51.5\%) and \texttt{dependency\_error} (24.6\%) together accounted for 76.1\% of all failures. Among the four recoverable types, \texttt{missing\_env} achieved the highest recovery rate at 61.29\%, while \texttt{syntax\_error}, the most frequent type, was recovered in 41.22\% of cases. Overall, the mechanism successfully recovered 156 out of 476 failures (32.77\%). 
The remaining 57 unrecoverable failures arose either when no diagnostic output was available, preventing identification of the failure cause, or when the error was due to environmental constraints beyond command-level repair, such as the LLM using incorrect file paths despite the environment specifications describing the correct ones.

The improvement varies considerably across reports, ranging from +3.20pp (TTPs\#11) to +33.77pp (TTPs\#8). To understand this variance, we analyzed failure-type distributions for four representative reports, as summarized in Table~\ref{tab:rq2-variance}.

\begin{table}[ht]
\centering
\caption{Failure distribution and recovery rate for
selected reports. Values: failure count (recovery rate \%)}
\label{tab:rq2-variance}
\scriptsize
\setlength{\tabcolsep}{4pt}
\begin{tabular}{|l|c|c|c|c|}
\hline
& \textbf{TTPs\#1} & \textbf{TTPs\#8} & \textbf{TTPs\#10} & \textbf{TTPs\#11} \\ \hline
\textbf{Improvement} & +18.09pp & +33.77pp & +6.11pp & +3.20pp \\ \hline
syntax\_error & 47 (23.4) & 16 (50.0) & 8 (75.0) & 16 (6.2) \\ \hline
dependency\_error & 13 (23.1) & 53 (41.5) & -- & 8 (0.0) \\ \hline
missing\_env & 12 (58.3) & 3 (100) & -- & 2 (100) \\ \hline
timeout & 5 (0.0) & 4 (50.0) & 6 (0.0) & 3 (0.0) \\ \hline
unrecoverable & 6 & 7 & 2 & 12 \\ \hline
\textbf{Total} & 83 & 83 & 16 & 41 \\ \hline
\end{tabular}
\end{table}

First, a high proportion of unrecoverable failures limits the recovery potential. TTPs\#11 had the highest unrecoverable ratio (29.3\%, 12 of 41 failures), and even its recoverable failures were rarely fixed (syntax\_error 6.2\%, dependency\_error 0\%), yielding the lowest improvement (+3.20pp). 

Second, reports with high initial success rates naturally show smaller improvements. TTPs\#7, TTPs\#9, and TTPs\#10 all had initial success rates above 80\%, leaving few failures for the mechanism to recover. Their improvements ranged from +6.11pp to +6.71pp, yet these reports still achieved final success rates between 88.47\% and 92.00\%.


Third, the distribution of root causes among failures affects recovery effectiveness. TTPs\#1 and TTPs\#8 both had 83 failures with similar initial success rates, yet their improvements differed substantially (+18.09pp vs.\ +33.77pp). In TTPs\#8, 53 dependency\_errors predominantly stemmed from a single recurring root cause---the LLM placed double backslashes where only one was needed in remote credential parameters---allowing it to apply a consistent fix across instances (41.5\% recovery). In TTPs\#1, 47 syntax\_errors arose from varying escaping contexts in web-shell commands, where attacker-issued commands were relayed through a web-shell and then reinterpreted by the underlying shell, yielding only 23.4\% recovery. Additionally, when the initial web-shell setup command itself failed in TTPs\#1, all subsequent commands that depended on it also failed, further limiting recovery opportunities.

\begin{table}[ht]
\centering
\caption{Attack Command Failure Type Classification and Recovery Examples}
\label{tab:command_fix_strategies}
\setlength{\tabcolsep}{1pt}
\scriptsize
\begin{tabular}{|l|p{4.8cm}|p{4.8cm}|} \hline
\textbf{Failure Type} & \textbf{Problem Description} & \textbf{Recovery Examples} \\
\hline

syntax\_error &
Escaping and string parsing errors &
\texttt{\textbackslash{}\textbackslash{} $\rightarrow$ \textbackslash{}}, \texttt{"\textbackslash{}"cmd\textbackslash{}"" $\rightarrow$ `"`cmd`"`} \\
\hline

syntax\_error &
Web shell invocation parameter conflict &
\texttt{cmd=raw command $\rightarrow$ cmd=\%20encoded\%20command} \\
\hline

syntax\_error &
Shell/binary compatibility issue &
\texttt{dir$\rightarrow$Get-ChildItem}, \texttt{del$\rightarrow$Remove-Item}, \texttt{copy$\rightarrow$Copy-Item} \\
\hline

syntax\_error &
HTTP upload format mismatch &
\texttt{-InFile $\rightarrow$ multipart/form-data body} \\
\hline

syntax\_error &
Large-file transfer failure &
\texttt{upload(file) $\rightarrow$ upload(file\_chunks[])} \\
\hline

syntax\_error &
Unstable output and redirection &
S\texttt{cmd \textgreater{} file $\rightarrow$ cmd \textbar{} Out-File file} \\
\hline

missing\_env &
Use of a restricted path &
\texttt{System32 $\rightarrow$ Users\textbackslash{}Public\textbackslash{}data} \\
\hline

missing\_env &
Failure to collect environment information &
\texttt{command failure $\rightarrow$ fallback output} \\
\hline

missing\_env &
Authentication and session ambiguity &
\texttt{admin $\rightarrow$ HOST\textbackslash{}admin} \\
\hline

missing\_env &
Missing execution preconditions &
\texttt{execute $\rightarrow$ prepare + execute} \\
\hline

dependency\_error &
Insufficient privileges (Access Denied) &
\texttt{cmd $\rightarrow$ RunAs(cmd)} \\
\hline

dependency\_error &
Remote access and authentication dependency &
\texttt{Invoke-Command $\rightarrow$ auth\_setup + Invoke-Command} \\
\hline

dependency\_error &
Missing modules or binaries &
\texttt{Get-ADComputer $\rightarrow$ nltest / net} \\
\hline

dependency\_error &
Resource conflict and duplicate execution &
\texttt{create $\rightarrow$ if exists then overwrite} \\
\hline

dependency\_error &
Network authentication bypass &
\texttt{\textbackslash{}\textbackslash{}host\textbackslash{}share $\rightarrow$ New-PSDrive + access} \\ \hline
\end{tabular}
\end{table}

Table~\ref{tab:command_fix_strategies} illustrates how the failure recovery mechanism addressed each failure type in practice. For \texttt{syntax\_error}, nested quotation marks were replaced with PowerShell backticks and CMD-based commands were rewritten as native PowerShell cmdlets. For \texttt{dependency\_error}, commands were wrapped with \texttt{Start-Process -Verb RunAs} or replaced with built-in alternatives (e.g., \texttt{nltest} instead of the Active Directory module). For \texttt{missing\_env}, file operations targeting protected paths were redirected to writable locations with preemptive directory creation.

\smallskip
\noindent\textbf{RQ3: What are the generated playbook size, time, and cost of the automated end-to-end adversary emulation?}

We measure three efficiency metrics for each report--model pair based on five runs: attack chain length (the number of Abilities in the generated playbook), time (playbook generation and failure recovery), and LLM API cost (calculated from total token usage). These metrics are compared against the manual authoring baseline: the ATT\&CK Evaluations Library~\cite{citd_eval} has released only 14 executable attack scenarios from 2019 to 2025, averaging approximately 5.1 months per scenario, and no prior studies on CTI-based adversary emulation~\cite{alam2023looking,choi2021probabilistic,ttpdrill,kumarasinghe2024semantic,li2022attackg,rahman2023attackers,extractor,takahashi2020aptgen} concretely report the time required for manual playbook construction.

\begin{table}[ht]
\centering
\caption{Attack Chain Length, Total Time, Token Usage, and Cost by Report(Claude Sonnet 4.5)}
\label{tab:rq1-main}
\scriptsize
\setlength{\tabcolsep}{1.2pt}
\begin{tabular}{|l|r|r|r|r|r|r|r|r|r|}
\hline
\textbf{ID} & \textbf{Len.} & \textbf{Total(s)} & \textbf{Gen.(s)} & \textbf{Mod.(s)} & \textbf{Exec.(s)} & \textbf{Prep.(s)} & \textbf{Input(K)} & \textbf{Output(K)} & \textbf{Cost(\$)} \\ \hline
TTPs\#1  & 33.6 & 1802.97 & 130.93 &  86.09 & 1398.20 & 187.75 & 69.4 & 13.5 & 0.41 \\ \hline
TTPs\#2  & 29.4 & 1065.22 & 255.45 &  16.31 &  670.60 & 122.86 & 73.3 & 17.1 & 0.48 \\ \hline
TTPs\#3  & 23.0 &  927.33 & 107.85 &  19.24 &  645.00 & 155.24 & 30.1 &  8.9 & 0.22 \\ \hline
TTPs\#4  & 30.4 & 1809.97 & 207.57 &  27.21 & 1394.40 & 180.79 & 53.8 & 14.2 & 0.38 \\ \hline
TTPs\#5  & 27.2 & 1679.44 & 230.97 &  38.57 & 1187.80 & 222.11 & 58.2 & 14.8 & 0.40 \\ \hline
TTPs\#6  & 24.2 &  790.28 & 128.84 &  21.14 &  485.40 & 154.90 & 36.2 &  9.9 & 0.26 \\ \hline
TTPs\#7  & 28.2 & 1452.84 & 166.86 &  23.11 & 1075.40 & 187.47 & 39.3 & 12.1 & 0.30 \\ \hline
TTPs\#8  & 33.8 & 2275.96 & 185.31 &  63.05 & 1806.00 & 221.60 & 67.2 & 17.8 & 0.47 \\ \hline
TTPs\#9  & 23.4 &  974.60 & 118.41 &  14.54 &  666.80 & 174.86 & 39.1 &  9.9 & 0.27 \\ \hline
TTPs\#10 & 22.6 & 1070.45 & 133.67 &   3.29 &  773.00 & 160.49 & 38.9 & 11.2 & 0.28 \\ \hline
TTPs\#11 & 24.0 & 1494.56 & 161.82 &  31.13 & 1092.60 & 209.01 & 65.6 & 13.1 & 0.39 \\ \hline
\textbf{Average} & \textbf{27.3} & \textbf{1394.87} & \textbf{166.15} & \textbf{31.24} & \textbf{1017.75} & \textbf{179.73} & \textbf{51.9} & \textbf{13.0} & \textbf{0.35} \\ \hline
\end{tabular}
\end{table}

As shown in Table~\ref{tab:rq1-main}, using Claude Sonnet 4.5, each playbook contained an average of 27.2 Abilities. Playbook generation required 2.8 minutes on average, and including failure recovery the total preparation time was approximately 3.3 minutes per report. The average LLM token cost was \$0.35 per report. These results suggest that the proposed method provides substantial time and cost savings compared with manual playbook authoring.

\medskip
\noindent\textbf{RQ4: How does the performance of the end-to-end adversary emulation system vary when different LLM models are used?}

\begin{figure}[ht]
\centering
\includegraphics[width=0.7\linewidth]{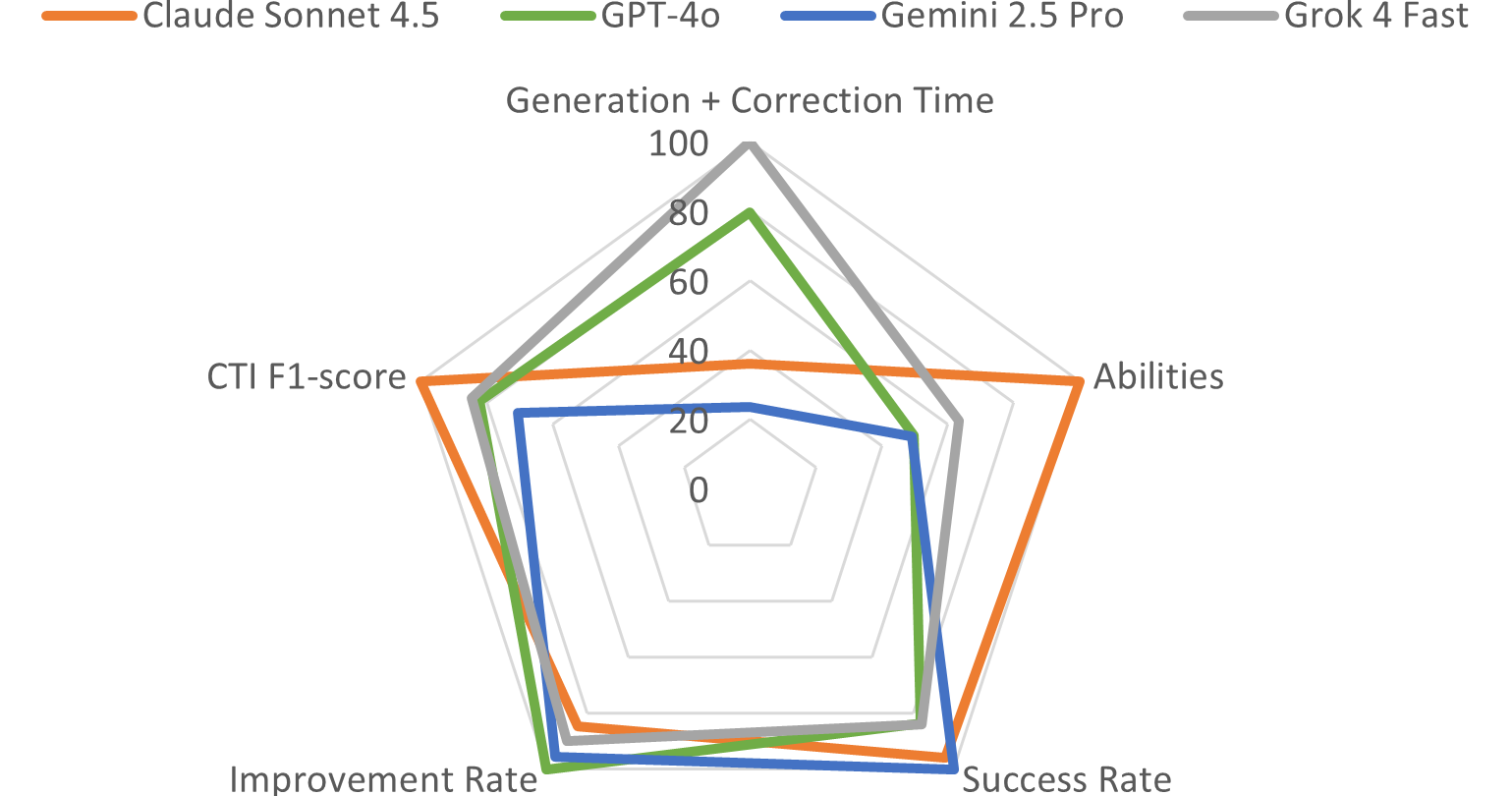}
\caption{Performance Comparison across LLM Models(0--100; best model for each metric = 100)}
\label{fig:rq5-radar}
\end{figure}

We compare four LLM models by aggregating the metrics from RQ1--RQ3. Figure~\ref{fig:rq5-radar} presents five key metrics: CTI fidelity, attack-chain length, execution success rate, improvement from failure recovery, and generation and revision time.

The results show that each model has a distinct area of strength.
%
Claude Sonnet 4.5 achieved the highest CTI F1-score (60.50\%) and the longest attack chains (27.2 Abilities), generating the most comprehensive and faithful playbooks among the four models. Its execution success rate of 84.22\% was also strong. Although its generation time (197.39s) was longer than GPT-4o and Grok 4 Fast, this overhead is modest compared with manual authoring and is justified by its superior CTI fidelity and playbook scale.
%
Grok 4 Fast was the most efficient model, requiring only 71.19 seconds and the lowest token cost, while maintaining a competitive CTI F1-score (51.10\%).
%
Gemini 2.5 Pro achieved the highest final execution success rate at 87.86\%, but this advantage was offset by the longest generation time (300.77s) and relatively short attack chains (13.3 Abilities), which resulted in the lowest CTI F1-score (42.58\%) among the four models.
%
GPT-4o showed the largest improvement from failure recovery (+17.23\,pp), indicating relatively strong recovery capability. Its CTI Precision was the highest at 77.51\%, while Recall remained low due to shorter attack chains.

Notably, the failure recovery mechanism yielded consistent gains of +14.59pp to +17.23pp across all four models, shifting execution success from an initial range of 56.33\%--71.37\% to a final range of 73.56\%--87.86\%, confirming that the mechanism is not dependent on a specific model but generalizes effectively across LLMs.

\section{Discussion}
\label{sec:discussion}

\subsection{Comparison with AURORA}

We first compare the datasets used in the two studies.
Table~\ref{tab:cti-report-stats} summarizes their basic statistics. While AURORA~\cite{aurora} used 250 reports averaging 5,561 words, the 11 CTI reports used in this study average 30,840 words, or about 5.5 times longer per report. This suggests that, although smaller in number, the dataset used in our study is sufficiently rich for evaluating adversary emulation from detailed CTI reports.

\begin{table}[ht]
\centering
\caption{Comparison of CTI report statistics between AURORA and this study}
\label{tab:cti-report-stats}
\setlength{\tabcolsep}{2pt}
\scriptsize
\begin{tabular}{|l|c|c|c|}
\hline
\textbf{Study} & 
\textbf{\# Reports} & 
\textbf{Words / Report} \\
\hline
AURORA~\cite{aurora} 
& 250 
& 5,561 (2,666--10,193) \\
\hline
This study 
& 11 
& 30,840 (15,896--59,104) \\
\hline
\end{tabular}
\end{table}

AURORA conducted its experiments using only a single model, GPT-o3. In contrast, our study performed a comparative evaluation using four recent models and found that each model exhibits distinct strengths.

For the comparative experiment, we used the same 10 CTI reports that AURORA used as ground truth in their evaluation.
As noted earlier, these reports explicitly contain Technique IDs in their main bodies, enabling objective measurement of Precision and Recall against report-derived ground truth.
We then used these 10 reports as input to our system, generated playbooks with Claude Sonnet 4.5, and executed them in the Windows environment provided by AURORA to enable a direct comparison of performance.

\begin{table}[ht]
\centering
\caption{Comparison with AURORA (Claude Sonnet 4.5, the 10 ground-truth CTI reports provided by AURORA).
\newline
\scriptsize\textit{Reports: (1) AA23-341A, (2) AA24-046A, (3) ALPHVBlackcat, (4) PhobosRansomware, (5) PlayRansomware, (6) RhysidaRansomware, (7) ScatteredSpider, (8) SnatchRansomware, (9) BlackBasta, (10) RoyalRansomware.}}
\label{tab:aurora-comparison}
\scriptsize
\setlength{\tabcolsep}{0.7pt}
\begin{tabular}{cccccccccccc}
\toprule
\multirow{2}{*}{ID} & \multicolumn{2}{c}{Chain Len.} & \multicolumn{3}{c}{Exec. Succ. (\%)} & \multicolumn{2}{c}{CTI Prec. (\%)} & \multicolumn{2}{c}{CTI Rec. (\%)} & \multicolumn{2}{c}{CTI F1 (\%)} \\
\cmidrule(lr){2-3} \cmidrule(lr){4-6} \cmidrule(lr){7-8} \cmidrule(lr){9-10} \cmidrule(lr){11-12}
& AURORA & Ours & AURORA & Ours\textsubscript{init} & Ours\textsubscript{final} & AURORA & Ours & AURORA & Ours & AURORA & Ours \\
\midrule
1 & 25 & 17 & 60.00 & 35.29 & 82.35 &  6.25 & 16.67 & 10.00 & 20.00 &  7.69 & 18.18 \\
2 & 25 & 13 & 64.00 & 53.85 & 53.85 & 20.00 & 25.00 & 27.27 & 18.18 & 23.08 & 21.05 \\
3 & 33 & 29 & 66.67 & 17.24 & 82.76 &  0.00 &  4.76 &  0.00 & 20.00 &  0.00 &  7.69 \\
4 & 26 & 28 & 61.54 & 35.71 & 78.57 & 82.35 & 52.63 & 40.00 & 28.57 & 53.85 & 37.04 \\
5 & 23 & 38 & 60.87 & 28.95 & 52.63 & 20.00 & 28.12 & 18.75 & 56.25 & 19.35 & 37.50 \\
6 & 27 & 41 & 48.15 & 51.22 & 63.41 & 37.50 & 46.15 & 30.00 & 60.00 & 33.33 & 52.17 \\
7 & 21 & 31 & 61.90 & 58.06 & 80.65 & 28.57 & 33.33 & 12.50 & 21.88 & 17.39 & 26.42 \\
8 & 21 & 36 & 66.67 & 13.89 & 66.67 & 50.00 & 25.00 & 38.89 & 33.33 & 43.75 & 28.57 \\
9 & 23 & 30 & 60.87 & 20.00 & 43.33 & 28.57 & 19.23 & 44.44 & 55.56 & 34.78 & 28.57 \\
10 & 23 & 27 & 56.52 & 14.81 & 51.85 & 14.29 & 22.22 & 16.67 & 33.33 & 15.39 & 26.66 \\
\midrule
\textbf{Avg.}  & \textbf{24.70} & \textbf{29.00} & \textbf{60.72} & \textbf{32.90} & \textbf{65.17} & \textbf{28.25} & \textbf{27.31} & \textbf{23.85} & \textbf{34.71} & \textbf{26.07} & \textbf{30.57} \\
\bottomrule
\end{tabular}
\end{table}

Table~\ref{tab:aurora-comparison} shows the results of evaluating both AURORA's publicly released executable attack chains and our generated playbooks under identical metrics: attack-chain length, execution success rate, CTI Precision, CTI Recall, and CTI F1-score.

The average chain length was slightly longer in our system, at 29.0, compared with 24.7 in AURORA, indicating a more elaborate multi-stage attack flow.

The experimental results showed that the average CTI F1-score was 30.57\% for our system and 26.07\% for AURORA, suggesting that the proposed method is sufficiently competitive overall. Although the ALPHVBlackcat report may appear to disproportionately affect this average—because its ground truth contains only five Technique IDs, yielding 0\% CTI Precision and Recall for AURORA but an F1-score of 7.69\% for our system—our system still outperformed AURORA in average CTI F1-score even after excluding this report.

In terms of final execution success rate, our system achieved 65.17\%, outperforming AURORA's 60.72\% by 4.45\%p. This improvement was primarily attributable to the failure recovery mechanism. Before recovery, our system achieved an initial execution success rate of 32.90\%, whereas AURORA reached 60.72\%. This gap reflects the contrasting command-generation strategies of the two systems: AURORA selects known-working commands from a pre-built library of 5,555 actions, whereas our system generates commands from scratch using an LLM. By revising failed commands and retrying execution, the recovery mechanism improved our success rate by 32.27\%p, allowing the final result to exceed that of AURORA, the state-of-the-art adversary emulation system.

Notably, the failure recovery mechanism achieved a larger improvement on the AURORA reports (+32.27\%p) than on our own dataset (+14.59\%p). This difference likely reflects the shorter and less detailed nature of AURORA's CTI reports (averaging 5,561 words vs. 30,840 in ours), which contain fewer attack details for the LLM to reference when generating executable commands, leading to more initial failures that the recovery mechanism can subsequently address.

Although this comparison was limited to the 10 ground-truth reports from AURORA's full dataset, the results show that our system achieves competitive CTI fidelity and, with the help of the failure recovery mechanism, higher execution success while additionally providing full end-to-end automation and failure-aware revision—capabilities that AURORA does not offer.

\subsection{Generating Multiple Playbooks from a Single CTI Report}
Recall that the current system generates only one playbook per CTI report. In Step 2, the LLM constructs a directed acyclic graph (DAG) whose nodes represent attack steps and whose edges represent dependencies. The system then derives a single execution order from this graph and converts it into one Adversary.
However, branching structures in the graph can produce multiple valid playbooks from a single CTI report. The number of such flows grows combinatorially as branch points increase. This makes exhaustive generation and evaluation impractical.

For this reason, the present study generates only one playbook per CTI report. We instead focused on building a fully automated pipeline and validating the effectiveness of the feedback-based failure recovery mechanism. Extending the system to generate and select among multiple playbooks is left for future work.

\section{Conclusion}
\label{sec:conclusion}
This paper proposed a \emph{fully end-to-end adversary emulation} system that supports all three key functions: playbook generation, execution, and feedback-based recovery from execution failures. To the best of our knowledge, this is the first work to provide complete end-to-end adversary emulation by integrating these three essential capabilities into a single pipeline. 
Our experiments on 11 high-quality CTI reports showed that the proposed system can generate executable playbooks while faithfully reflecting the attack techniques described in the original CTI reports. The results also showed that both the time and cost required for playbook generation and revision remain low, demonstrating practical efficiency compared with manual adversary emulation. In particular, the failure recovery mechanism consistently improved execution success rates across different failure types, thereby increasing the practical usability of the generated playbooks.

The current system assumes that the test environment and the environment specification are prepared in advance by the security operator. These steps could be further automated by integrating Infrastructure as Code tools such as \textit{Terraform}~\cite{terraform} for environment provisioning and system information collectors such as \textit{osquery}~\cite{osquery} or autonomous AI agents such as \textit{OpenClaw}~\cite{openclaw} for environment specification generation. Extending the automation to cover these remaining steps is a key direction for future work.

\appendix
\section{Appendix}
This appendix includes examples of the system inputs, the intermediate outputs produced at each pipeline step, and the LLM prompts.

\medskip
\noindent
Listing~\ref{lst:cti-example} shows an excerpt from the KISA TTPs\#1 report, \textit{Compromising an Internal Network through a Website}. The report is structured by first listing tactics such as \textit{Initial Access}, and then enumerating techniques under each tactic, such as \textit{Valid Accounts}.

\noindent
\begin{minipage}{\linewidth}
\begin{lstlisting}[language={}, caption={Excerpt of TTPs\#1 CTI report}, basicstyle=\scriptsize\ttfamily, breaklines=true, breakatwhitespace=false, label={lst:cti-example}]
# Initial Access
1. Valid Accounts
Successful login to the corporate website using previously collected valid account credentials.

2019-06-07 02:40:07 211.115.xxx.xxx GET /default.asp islogin=y&stf_id=xxx&stf_name=
stf_cId=xxx&stf_ib=xxx&stf_IdF=xxx&stf_cName=[]&stf_usertype=&stf_userlevel=&stf_fleCode=&stf_woalcode=0&id_cookie=80 - 61.254.xxx.xxx

### Mitigation Strategies
- Restrict user access by IP address.
- Monitor web logs to detect access attempts from unauthorized IP addresses.
\end{lstlisting}
\end{minipage}

\vspace{30pt}

\noindent
Listing~\ref{lst:env-spec} shows an excerpt from the environment specification for TTPs\#1.

\noindent
\begin{minipage}{\linewidth}
\begin{lstlisting}[language={}, caption={Excerpt of Environment Specification File (TTPs\#1)}, basicstyle=\scriptsize\ttfamily, breaklines=true, breakatwhitespace=false, label={lst:env-spec}]
## Environment Information
- Shell: Windows PowerShell 5.1

## Attacker Server (192.168.56.1:34444)
- GET /agents/* : Download files required for the victim's PC
    - sandcat_ttps1.ps1 : Caldera agent (Executes after initial privilege escalation)
    - cmd.asp : Web shell file (Executed via cmd.asp?cmd={command})
    - PrintSpoofer64.exe : Privilege escalation tool (Runs with IIS AppPool privileges)
- POST /upload : Endpoint for exfiltrating victim data

## IIS Web Application (192.168.56.105)
- Login: POST http://192.168.56.105/login_process.asp
    - Credentials: admin / P@ssw0rd!2020
- File Upload: POST http://192.168.56.105/upload_handler.asp
    - Form field: file, Upload path: /uploads/

## Victim Server (192.168.56.105)
- Target files for exfiltration: C:\Users\Public\data\*
- Additional info collection: Search only under the "Users" folder and save to the path
\end{lstlisting}
\end{minipage}

\newpage

\noindent
The LLM prompt used to extract these abstract attack goals from the report is shown in Listing \ref{lst:step1-chunk-prompt}.

\noindent
\begin{minipage}{\linewidth}
\begin{lstlisting}[basicstyle=\scriptsize\ttfamily, breaklines=true, breakatwhitespace=false, caption={Abstract Attack Flow Generation Prompt for Step 1}, label={lst:step1-chunk-prompt}]
You are analyzing a KISA threat intelligence report chunk-by-chunk to extract 
attack goals.

# Overview
{overview}

{previous_context}
# Current Chunk ({chunk_num}/{total_chunks})
{chunk}

# Task
Analyze this chunk and identify any NEW attack goals not yet found.

Focus on:
- Goals NOT already in the "Previously Identified Goals" list
- Environment-independent objectives (no IPs, URLs, credentials)
- MITRE ATT&CK tactics: initial-access, execution, persistence, privilege-escalation, defense-evasion, credential-access, discovery, lateral-movement, collection, exfiltration, impact

## Output Format (JSON only)
{
  "new_goals": [
    {
      "goal": "Clear description of attack objective",
      "tactic": "MITRE ATT&CK tactic name",
      "description": "Brief explanation"
    }
  ],
  "report_complete": true/false
}

Output JSON only. No explanations.
\end{lstlisting}
\end{minipage}

\newpage

\noindent
Listing \ref{lst:step2-prompt} presents the prompt used for generating specific commands and mapping MITRE ATT\&CK techniques to the target environment.

\noindent
\begin{minipage}{\linewidth}
\begin{lstlisting}[basicstyle=\scriptsize\ttfamily, breaklines=true, breakatwhitespace=false, caption={Concrete Attack Flow Generation Prompt for Step 2}, label={lst:step2-prompt}]
You are a penetration testing expert with deep knowledge of MITRE ATT&CK.

# Abstract Attack Goals
{abstract_flow}

# Target Environment
{environment_description}

# Task
Map each abstract goal to concrete attack steps using environment details.
For each step, select the most appropriate MITRE ATT&CK technique.

## Core Principles
1) Every node MUST include `environment_specific.commands`
   - Single-line command for target shell; chain with `;`
   - No code blocks/comments; plain command string only
   - Add simulation stub (e.g., echo) if real command impossible

2) Enforce environment detail usage
   - Reflect IP/URL/credentials/params from environment in commands
   - No external downloads unless explicitly provided

3) Tool priority: built-in OS tools > simulation command

4) Preserve all environment-specified values (params, credentials, URLs)

5) MITRE ATT&CK Technique Selection
   - Select ONE valid technique ID per node (e.g., T1059.001, T1190)
   - Prefer sub-techniques over parent techniques when applicable
   - Technique MUST be appropriate for {os_type} platform

## Output Structure (use YAML literal block scalar | for commands)
nodes:
  - id: "node_001"
    name: "Attack step name"
    tactic: "initial-access"
    technique:
      id: "T1190"
      name: "Exploit Public-Facing Application"
    environment_specific:
      target: "actual IP"
      url: "actual URL"
      credentials: {username: "user", password: "pass"}
      commands: |
        Single-line command for target shell

edges:
  - {from: "node_001", to: "node_002", dependency_type: "required"}

execution_order: ["node_001", "node_002"]

Output YAML only. No explanations.
\end{lstlisting}
\end{minipage}

\newpage

\noindent
Listing \ref{lst:step4} provides an excerpt of the generated Caldera Ability and Adversary.

\noindent
\begin{minipage}{\linewidth}
\begin{lstlisting}[language={}, caption={Caldera Ability and an Adversary Excerpted from Step 3 Output (TTPs\#1)}, basicstyle=\scriptsize\ttfamily, breaklines=true, breakatwhitespace=false, label={lst:step4}]
[abilities.yaml]
- ability_id: 8497abc5-e5c1-57c0-af6f-cded271c90d6
name: Upload web shell to bulletin board
description: Exploit file upload vulnerability to deploy web shell
tactic: initial-access
technique_id: T1190
technique_name: Exploit Public-Facing Application
executors:
    - name: psh
    platform: windows
    command: '$wc = New-Object System.Net.WebClient;
    $wc.UploadFile("http://192.168.56.105/upload_handler.asp",
    "C:\\Users\\Public\\data\\cmd.asp")'
    timeout: 60
- ability_id: cce02e18-917a-5a8b-a070-8857fa81d541
...

[adversaries.yaml]
- adversary_id: kisa-ttp-adversary-KISA_TTPs_1
  name: KISA TTP Adversary (KISA_TTPs_1)
  description: Auto-generated adversary profile from KISA TTP report
  atomic_ordering:
  ...
  - 8497abc5-e5c1-57c0-af6f-cded271c90d6
  - cce02e18-917a-5a8b-a070-8857fa81d541
  ...
\end{lstlisting}
\end{minipage}

\vspace{5pt}

\noindent
Listing \ref{lst:recovery-prompt} details the prompt used to guide the LLM in fixing failed abilities during this recovery phase.

\noindent
\begin{minipage}{\linewidth}
\begin{lstlisting}[basicstyle=\scriptsize\ttfamily, breaklines=true, breakatwhitespace=false, caption={Command Revision Prompt for the failure recovery layer}, label={lst:recovery-prompt}]
You are an expert in fixing Caldera Ability commands.
Analyze the failed Ability and fix the command.

[Failed Ability]
ID: {ability_id}, Name: {ability_name}
Tactic: {tactic}, Technique: {technique_id} ({technique_name})

[Original Command]
{original_cmd}

[Execution Result - Failed]
Exit Code: {exit_code}
stderr: {stderr}
stdout: {stdout}

[Failure Type: {failure_type}]
{failure_type_description}

[Environment Description]
{env_description}

{correction_history}

[Important Rules]
1. Each Ability runs independently - no variable sharing
2. Use actual values from Environment Description
3. Follow shell constraints in Environment Description
4. Output only the fixed command (no explanation)

Generate the fixed command:
\end{lstlisting}
\end{minipage}

\end{document}